\documentstyle[emulateapj]{article}

\def\lesssim{\mathrel{\hbox{\rlap{\hbox{\lower4pt\hbox{$\sim$}}}\hbox{$<$}}}}
\def\gtrsim{\mathrel{\hbox{\rlap{\hbox{\lower4pt\hbox{$\sim$}}}\hbox{$>$}}}}

\begin{document}
\submitted{December 18th 2002}
\title{A Joint Sunyaev-Zel'dovich Effect and X-ray Analysis of Abell
3667}

\author{C.M. Cantalupo\altaffilmark{1}, A.K. Romer, J.B. Peterson, P. Gomez,
        G. Griffin, M. Newcomb, R.C. Nichol}
\affil{Department of Physics, Carnegie Mellon University, 5000. Forbes Ave.,
Pittsburgh, PA--15213}

\altaffiltext{1}{Current affiliation: University of California
Berkeley Space Sciences Laboratory}

\begin{abstract} 

We present a 40GHz (7.5 mm) raster scan image of a
$3.6\arcdeg\times2\arcdeg$ region centered on the low redshift
($z=0.055$) cluster of galaxies Abell 3667. The cluster was observed
during the Antarctic winter of 1999 using the Corona instrument
($15.7\arcmin$ FWHM beam) on the Viper Telescope at the South Pole.
The Corona image of A3667 is one of the first direct ({\it i.e.}
rather than interferometer) thermal Sunyaev-Zel'dovich effect images
of a low redshift cluster. The brightness temperature decrement at the
X-ray centroid ($20^h 12^m 28.9^s, -56\arcdeg 49\arcmin 51\arcsec$
J2000) was measured to be $\Delta T_{\rm CMB}=-154\mu K$.  We have
used the 40GHz map of A3667 in conjunction with a deep ROSAT PSPC
(X-ray) image of the cluster, to make a measurement of the Hubble
Constant. We find $H_0= 64^{+96}_{-30}$ km s$^{-1}$ Mpc$^{-1}$ (68\%
confidence interval).  Our $H_0$ calculation assumes that the cluster
can be described using an isothermal, tri-axial ellipsoidal,
$\beta$-model and includes several new analysis techniques including
an automated method to remove point sources from X-ray images with
variable point spread functions, and an efficient method for
determining the errors in multi-parameter maximum likelihood analyzes.
The large errors on the $H_0$ measurement are primarily due to the
statistical noise in the Corona image. We plan to increase the
precision of our measurement by including additional clusters in our
analysis and by increasing the sensitivity of the Viper SZE maps.

\end{abstract}

\section{Introduction}
\label{sec:intro}

The Sunyaev-Zel'dovich effect (Sunyaev \& Zel'dovich 1972, SZE
hereafter) describes the inverse Compton scattering of cosmic
microwave background (CMB) photons by energetic free electrons.  For
instance, the random thermal motions of electrons trapped in the
potential wells of clusters of galaxies result in a frequency
dependent change in CMB intensity known as the thermal SZE.
Similarly, the bulk peculiar motion of these electrons results in
frequency independent change known as the kinetic SZE.  Observations
of the thermal and kinetic SZE using either single dish telescopes
({\it e.g.} Mason, Myers \& Readhead 2001; Pointecouteau et al. 2001
\& 2002; De Petris et al. 2002 or interferometric techniques ({\it
e.g.} Jones et al. 2001; Reese et al. 2001 \& 2002; Udomprasert, Mason
\& Readhead 2001) have improved dramatically in recent years (see
Birkinshaw 1999 for a recent review).  Measurements of the SZE have
been used for a variety of scientific applications, including the
estimation of the Hubble Constant (e.g. Mason, Myers, \& Readhead
2001; Jones et al. 2001; Reese et al. 2002); determinations of the
fraction, by mass, of baryons in the Universe (e.g. Grego et
al. 2001); X-ray independent measurements of cluster temperatures
(e.g. Pointecouteau et al. 2002); and constraints on cluster peculiar
velocities (Holzapfel et al. 1997b, LaRoque et al. 2002).

The technique of interferometric SZE imaging is now well
established. By contrast, only recently has it become possible to
generate direct (rather than aperture synthesis) SZE images using
single dish telescopes. To date, only two clusters, RX J1347-1145 \&
RX J2228+2037, both at $z\simeq0.4$, have published direct SZE images
(Pointecouteau et al. 1999, 2001 \& 2002; Komatsu et al.  2001). We
report here a third direct SZE image. This 40 GHz (7.5 mm) image of
the ($z=0.055$, Sodre et al. 1992) Abell cluster (Abell, Corwin, \&
Olowin 1989) A3667 was made using the Viper telescope at the South
Pole as part of the Viper Sunyaev-Zel'dovich Survey (VSZS). The VSZS
aims to study a complete sample of southern clusters at
radio/microwave, optical and X-ray wavelengths and it goals include
the measurement of the Hubble Constant.  We chose to study A3667 as
our initial VSZS target because, as one of the brightest X-ray
($L_{0.5-2.0 {\rm keV}} = 4.1\times10^{44}$ ergs $s^{-1}$, David et
al. 1999) clusters in the REFLEX catalog (B\"ohringer et al. 2001),
A3667 is expected to have a strong SZE signal. A3667 is also one of
the best studied clusters in the sky. Supporting data at other
wavelengths include: X-ray imaging and spectroscopy (ROSAT: Knopp,
Henry \& Briel 1996; ASCA: Markevitch, Sarazin, Vikhlinin 1999; White
et al. 2000; BeppoSAX: Fusco-Femiano et al. 2001; Chandra: Vikhlinin,
Markevitch \& Murray 2001 a\& b; XMM), optical data in the form of
images, weak lensing mass maps and multi-object spectroscopy ({\it
e.g.}  Joffre et al. 2000; Katgert et al. 1998), and radio maps
(R\"ottgering et al.  1997; Hunstead et al. 1999).  In addition, A3667
has been the focus of three-dimensional MDH/N-body numerical
simulations (Roettiger, Burns \& Stone 1999).


An outline of the paper is as follows. In section~\ref{sec:viper-data}
we review our observing strategy and data reduction methods. We also
present the Corona map of the area around A3667. In
section~\ref{sec:Ho} we describe a joint fit, to an isothermal,
tri-axial ellipsoidal $\beta$-model, to the ROSAT Position Sensitive
Proportional Counter (PSPC) and Corona images of A3667 and present an
estimate of the Hubble Constant. In section~\ref{sec:conclusions}, we
present conclusions and discuss future plans for cluster observations
with Viper.


\section{The Corona Observation of A3667}
\label{sec:viper-data}

The observations presented herein were made with the 2.15 meter Viper
telescope during the austral winter of 1999. At the time, Viper was
fitted with a two pixel receiver instrument known as Corona that
measured the total incident power in the frequency range 38 to 44 GHz
($\bar\nu=40$, Coble et al.  1999) using HEMT (high electron mobility
transistor) amplifiers. The following features make Viper ideally
suited to the observation of low redshift clusters such as A3667; its
location at the South Pole, the two story conical shield and the flat
chopping tertiary mirror that sweeps the telescope beam backward and
forward by up to $3.6\arcdeg$ in co-elevation. The South Pole is one
of the driest sites in the world (Peterson et al. 2002), the shield
minimizes ground pick up and the chopper permits observations of
clusters that are extended over large areas of the sky.

Using Corona on Viper, we have generated a 40 GHz image of a
$2\arcdeg\times3.6\arcdeg$ region centered on A3667. We did so using a
raster scan technique that involved taking measurements during eight
complete chopper cycles at each of 25 fixed elevation positions
separated by $300\arcsec$ (roughly one third of a beam). For the A3667
observation, we used a $3.6\arcdeg$ chopper throw (the maximum
possible) and read out the instrument 512 times per 2.15 s$^{-1}$
chopper cycle, resulting in measurements separated by $50.6\arcsec$ in
the azimuth direction. Including overheads, each raster scan map took
a little over a minute to complete. In total, 3,103 raster scans were
made over six contiguous days: May 28, 29, 30, 31 \& June 1, 2 1999.
Every two hours during that period, the Carina nebula was observed as
a pointing check.  During our observation, the Corona beam pattern was
well described by a radially symmetric Gaussian with a FWHM of
$0.261\arcdeg$ ($15.7\arcmin$) and the total calibration uncertainty
was 8\% (Peterson et al 2000).

The raster scan data were initially stored as a time stream of
detector voltages.  The first stage of the analysis was to apply a
filter to the time stream. This corrected for the detector response
function; removed high frequency noise; and eliminated any signal
coupled to the power mains, by removing harmonics of 60 Hz.  Next
noisy and/or poor quality scans were flagged and cut from the
analysis: Those with any $\geq 5\sigma$ voltage spikes, those with
tracking or chopping position errors greater than two arc-minutes,
those for which the amplifiers were warmer than 25K, those with
excessive noise at harmonics of 60 Hz, those with rapidly varying
voltages and those with an unusually high average voltage.  We note
that the first data cut removed electrical interference in the data,
and the latter two helped remove time periods when clouds were moving
across the field of view.  Once all seven cuts had been applied, 1518
scans of the initial 3103 remained or $\simeq$36 hours of integration.

As it is conventional to express CMB intensity changes as changes in
brightness temperature, $\Delta T$, the next stage of the analysis
involved converting the voltage time stream into a $\Delta T$ time
stream using the detector calibration. Then the 1518 sets of time
stream were converted into grid maps consisting of 512 rows and 25
columns each. Next, the grid maps needed to be corrected for
contaminating signals known collectively as ``chopper synchronous
offsets''.  These offsets can include signals originating from the
ground, the atmosphere and the telescope components. The offset level
varies with chopper position because, as the chopper moves, the focal
plane is illuminated by different areas of the telescope optics (see
Peterson et al. 2000). Under the assumption that the offsets change
linearly with elevation, and maintained the same general waveform in
the azimuth (chopping) direction, the offsets were removed from the
inner regions of the scans as follows.  The linear component in the
elevation direction was removed by fitting to the top and bottom
4 rows in each of the columns.  Next, the linear component in the
azimuth direction was removed by fitting to the first and last 50
columns in each of the rows. After offset correction, the
$512\times25$ points in the offset subtracted grid maps were assigned
sky coordinates using the Peterson et al. (2000) telescope pointing
model. Finally, the maps were converted into fully sampled images with
$113\arcsec\times113\arcsec$ pixels. Linear interpolation in the
elevation direction was used to create these images because the grid
points in the raster scan were separated by $300\arcsec$ in elevation
(compared to $51\arcsec$ in azimuth).

\begin{figure*}
\vspace{-3.in}
\hspace{0.5in}{\centering\leavevmode\epsfysize=8.in \epsfbox{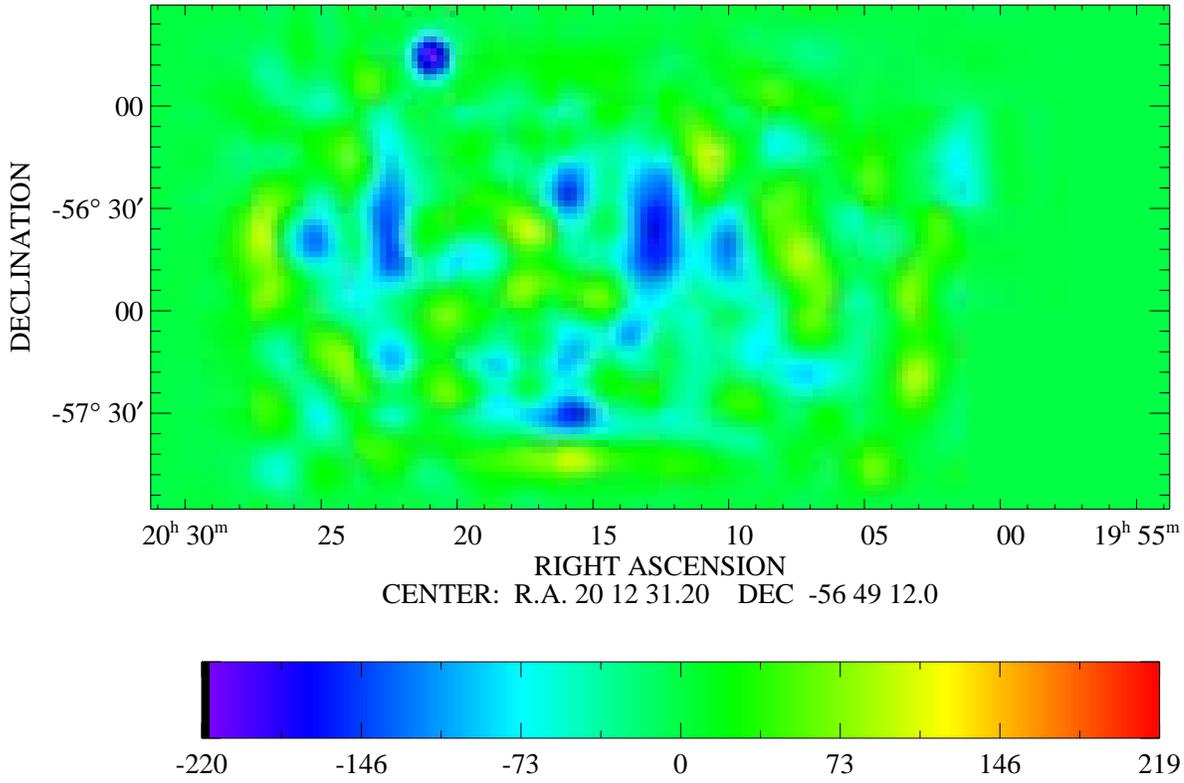}}
\vspace{-2.in}

\figcaption{\label{Fig:SZall} The $3.6\arcdeg\times2\arcdeg$ area of
the Corona 40 GHz map of A3667 in units of $\mu K$. The map is a
result of averaging 1518 raster scans of the region, for a total
exposure time of $\simeq 36$ hours. The map has been smoothed with a
$10.3\arcmin$ FWHM Gaussian kernel, which gives it an effective
resolution of $18\arcmin$ FWHM (the beam size of the instrument is
$15.7\arcmin$). The pixel size is $113\arcsec\times113\arcsec$.  The
large ($\simeq 30\arcmin$ in length), cold (blue), area in the center
of the map coincides with the X-ray emission from the cluster.}
\end{figure*}

\begin{figure*}

\hspace{0.75in}{\centering\leavevmode \epsfysize=8.in \epsfbox{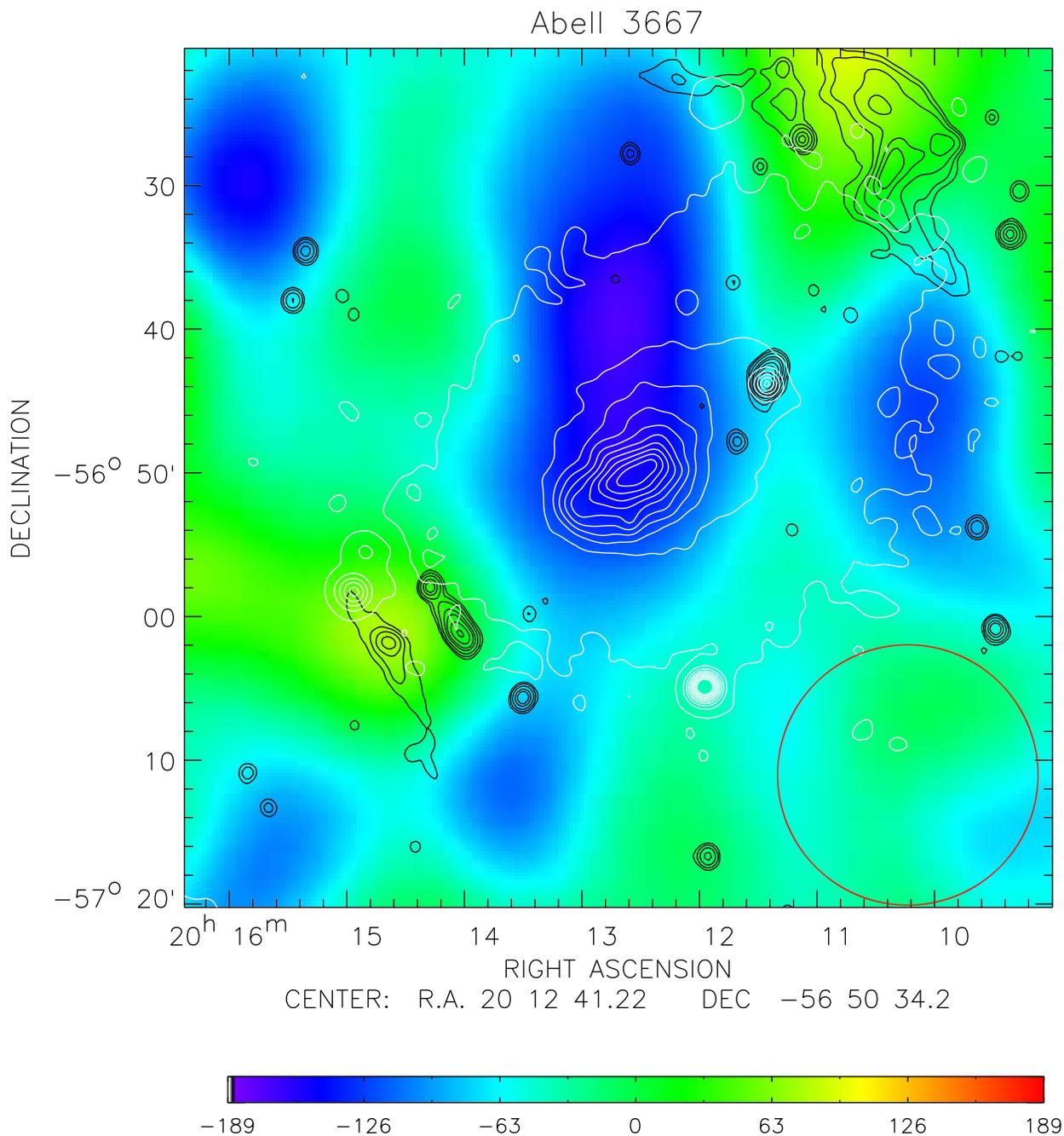}}
\figcaption{\label{Fig:SZ1degree} The central $1\times1\arcdeg$ region
of the Corona map in units of $\mu K$. The Corona have been re-binned
to the ROSAT PSPC pixel scale ($15\arcsec\times15\arcsec$). The black
contours represent the MOST radio (843 GHz) observation of A3667
(Hunstead, private communication). We note the correspondence between
the extended radio halo and hot spots in the Corona map. The white
contours represent the ROSAT PSPC 0.4-2.0 keV image of A3667. The
diameter of the red circle in the bottom right corner shows the
effective beam ($18\arcmin$ FWHM) of the Corona map. }
\end{figure*}

In Figures~\ref{Fig:SZall} and \ref{Fig:SZ1degree} we present the
result of averaging the 1,518 fully sampled images together. In these
Figures, the data have been smoothed with a $10.3\arcmin$ FWHM
Gaussian kernel, which gives them an effective resolution of
$18\arcmin$ FWHM (the beam size of the instrument was $15.7\arcmin$).
The effective beam FWHM has been represented by the red circle in the
bottom right corner of Figure~\ref{Fig:SZ1degree}. In
Figure~\ref{Fig:SZall} we present the full dimensions of the co-added
image, including the edge regions used for the offset subtraction. The
elongated blue (cold) region in the center of the image corresponds to
the position of A3667. Other cold structures in the map most likely
correspond to fluctuations due to receiver noise, primary CMB
anisotropies, or artifacts of the offset removal.  It is unlikely,
given the sensitivity and spatial resolution, that any of the cold
spots in the map correspond to serendipitous cluster detections.  Over
the inner region of the smoothed image, {\it i.e.}  the region not
used for offset subtraction, the standard deviation per
$113\arcsec\times 113\arcsec$ pixel was found to be $\simeq 60 \mu$
K. When deriving this random noise estimate, we assumed that the noise
was Gaussian and that each scan was independent. We tested the latter
assumption by re-calculating the noise after averaging into groups of
first five and then twenty consecutive scans. Grouping the scans in
this way before calculating the standard deviation in each pixel did
not change the noise estimate. This supports our assumption that the
noise was uncorrelated in time; if it had been, this test would have
resulted in an increase in the measured standard deviation.

In Figure~\ref{Fig:SZ1degree} we compare the inner
$1\arcdeg\times1\arcdeg$ of the Corona map with X-ray and radio images
of the same region. Note that for these comparisons, the Corona data
were rebinned to the PSPC pixel scale ($15\arcsec\times 15\arcsec$)
and that this smooths out the ($113\arcsec\times 113\arcsec$)
pixelization apparent in Figure~\ref{Fig:SZall}.  The white contours in
the Figure correspond to the 0.4-2.0 keV X-ray count rate as
measured by the PSPC instrument on the ROSAT satellite. The X-ray
count rate map was generated from data in the ROSAT public archive
using the data reduction pipeline designed for the SHARC Survey (Romer
et al. 2000). This data reduction pipeline was based on the ESAS
routines of Snowden et al. (1994) and includes the production of a
complementary 0.4-2.0 keV exposure map that was used during the
model fitting described below (Section~\ref{subsec:max-likefit}).  The
black contours correspond to the interferometric radio map at 843 MHz
created with the MOST interferometer (Hunstead et al. 1999; Richard
Hunstead, private communication). 

From Figure~\ref{Fig:SZ1degree}, it is clear that there is a
significant extended cold area close to the center of the X-ray
emission. At the centroid ($20^h 12^m 28.9^s, -56\arcdeg 49\arcmin
51\arcsec$ J2000) of the X-ray emission, the temperature decrement in
the smoothed Corona map is $\Delta T_{\rm CMB}=-154\mu $K.  There is
also a some evidence for correlations between hot regions in the image
and the MOST radio contours.  Taken together, the various correlations
between the Corona map and the X-ray and radio maps support the claim
in Peterson et al. (2000) that the Corona pointing reconstruction is
accurate to within $\pm 4\arcmin$. The extended radio emission to the
north-west of the cluster has been discussed in R\"ottgering et
al. (1997) and Roettiger, Burns and Stone (1999). Both groups suggest
that a merging event has produced turbulence in the outer region of
the cluster which in turn amplifies magneitc fields, reaccelerates
relativistic particles and thus produces radio emission. Radio
contamination can be a serious concern when creating SZE images, {\it
e.g.} RX J1347-1145 (Pointecouteau et al. 1999, 2001; Komatsu et al.
2001). However, as there is no evidence of radio contamination in the
region of the cluster used for the $H_0$ measurement
(Section~\ref{sec:Ho}), {\it i.e.} the core of the cluster, we do not
discuss the radio emission from A3667 again in this paper.

As mentioned above, the random noise per pixel in
Figure~\ref{Fig:SZall} is $\simeq 60 \mu$ K. In addition, the images
are contaminated by primary CMB anisotropies.  We have estimated the
level of the CMB contamination in the Corona images to be $\simeq38
\mu$K RMS using the Netterfield et al. (2001) anisotropy power
spectrum and assuming a top hat window function with
$10<l<400$. Therefore, some of the features visible in this Figure
could correspond to $>2\sigma$ CMB fluctuations. For this reason, we
cannot rule out that the signal at the position of A3667 is free from
CMB contamination.  The northern extension of the central cold region
could be due to such contamination. Recent observations of A3667 by
the multi-frequency Arcminute Cosmology Bolometer Array Receiver
(ACBAR, Runyan et al. 2002) on Viper should allow us to quantify the
level of CMB contamination in the Corona map because ACBAR has
detectors sensitive to frequencies above and below the
null\footnote{The thermal SZE results in a temperature decrement
(increment) below (above) $\nu=220$ GHz. At the cross over, or null,
frequency there is no net change in CMB brightness temperature.} in
the thermal SZE spectrum.

\section{Derivation of the Hubble Constant}
\label{sec:Ho}

It has been known for some time ({e.g.} Silk \& White 1978; Cavaliere,
Danese \& Zotti 1979) that a measurement of the thermal SZE toward a
cluster could result in a measurement of the Hubble Constant ($H_0$)
when it is combined with X-ray observations.  This is because the the
SZE is proportional to the line of sight integral of $T_e n_e$,
whereas the X-ray surface brightness is proportional to the line of
sight integral of $T_e^{\frac{1}{2}} n_e^2$.  These differing
dependencies on the electron temperature and number density, allow one
to solve for $H_0$, assuming we can find an accurate model for the
distribution of these parameters along the line of sight.

We have used the A3667 data described above to constrain $H_0$. We
have performed a joint maximum likelihood
(Section~\ref{subsec:max-likefit}) analysis of the Corona and ROSAT
PSPC images. During this analysis we have generating many thousands of
pairs of simulated ROSAT and Corona A3667 images by making various
assumptions concerning both the three dimensional distribution of
electrons in the cluster and the performance and noise characteristics
of the ROSAT PSPC and Viper Corona instruments (\S~\ref{sec:sim-pspc}
and \ref{sec:sim-viper}). The technique follows that used by other
authors {\it e.g.}  Reese et al. (2002) and Grainge et al. (2002), but
is the first application of the method to real space, rather than
Fourier space, analysis.

The assumption underlying our analysis was that the number density of
electrons in A3667 follows an isothermal $\beta$, or King (King 1962),
model.  The $\beta$-model can be parameterized by:

\begin{equation} 
n_e =n_{e0}\left( 1 + \left(\frac{r}{r_c}\right)^2\right)^{-3\beta/2},
\label{eqn:beta-ne} \end{equation}

where $n_{e0}$ is the central electron number density; $r$ is the
radius from the center of the cluster; and $r_c$ is the core radius
(Cavaliere \& Fusco-Femiano 1978). The X-ray emission from A3667 is
clearly not spherically symmetric (Figure~\ref{Fig:SZ1degree}), so we
have further assumed that the cluster has a tri-axial ellipsoidal
geometry (Zaroubi et al. 2001 have shown that the ellipsoidal $\beta$
model provides a good approximation to the three dimensional structure
of simulated clusters). We define the radius in
Equation~\ref{eqn:beta-ne} as follows:

\begin{equation} 
r^2 = x^2 + (\varepsilon y)^2 + (\varepsilon^\zeta l)^2, 
\label{eqn:ellipse} \end{equation} 

where the $x, y$, and $l$ directions are measured along the principle
axes of the ellipsoid. The axes are defined so that $x$ and $y$ are in
the plane of the sky and $l$ lies along the line of sight. There are
two isomorphic sets of models, those with $\varepsilon<1$ and those
with $\varepsilon>1$. For the model fitting described below
(Section~\ref{subsec:max-likefit}), we used $\varepsilon<1$, {\it
i.e.} the ellipticity was given by the ratio of the minor to major
axes. With this convention, the $x$ axis lies along the minor axis and
the $y$ axis along the major axis. We have parameterized the stretch
along the $l$ axis by $\varepsilon^\zeta$. If $\zeta=0$, then
$\varepsilon^\zeta=1$, and the stretch along the $l$ axis is equal to
that along the $x$ and the ellipsoid is described as prolate. If
$\zeta=1$, then $\varepsilon^\zeta=\varepsilon$, and the stretch along
the $l$ axis is equal to that along the $y$ axis and the ellipsoid is
described as oblate.

We have chosen to use the ellipsoidal $\beta$ model with $\zeta=0.5$
for our analysis because, as shown by Grainger (2001), this particular
model does not result in a significant bias in the measured $H_0$
value when applied to an ensemble of clusters taken from N-body
simulations.  The $\beta$ model is physically motivated; it provides a
reasonable, although not exact (see Sarazin 1988), description of the
radial dependence of the electron density in isolated spheres of
isothermal gas in hydrostatic equilibrium. However, it does have
several inadequacies. For example it predicts there will be a finite
electron number density even at very large radii from the cluster
center. It also includes an assumption of isothermality when certain
clusters, including A3667, are clearly not isothermal (Vikhlinin,
Markevitch \& Murray 2001a). Moreover, many clusters, including A3667,
deviate from spatial symmetry. For these reasons, future analyzes of
SZE images would benefit from the use of non-parametric MDM-N-body
simulations that have been constrained by observations rather than
generic $\beta$ models.

\subsection{Simulated ROSAT PSPC Images}
\label{sec:sim-pspc}

For a cluster, the X-ray surface brightness is proportional to the
line of sight integral of the square of the electron number
density. Assuming $n_e$ follows Equation~\ref{eqn:beta-ne}, using
infinite integration limits, and expressing surface brightness as a
count rate, $C_{\rm X}$, one obtains;

\begin{equation} C_{\rm X}(\theta_r) = C_{{\rm X} 0}\left(1
+\frac{\theta_r^2}{\theta_c^2}\right)^{-3\beta+1/2},
\label{eqn:xray-sb} 
\end{equation} 

where $C_{{\rm X} 0}$ is the central surface brightness and $\theta_c$
and $\theta_r$ are related to $r_c$ in Equation~\ref{eqn:beta-ne} and
to $r$ in Equation~\ref{eqn:ellipse}, respectively, as follows:
$\theta_c=r_c/D_A$ and $\theta_r=r_{l=0}/D_A$, where $D_A$ is the
angular diameter distance and $r_{l=0}$ is given by
Equation~\ref{eqn:ellipse} when $l=0$.

Using Equations~\ref{eqn:ellipse} and \ref{eqn:xray-sb}, we were able
to construct simulated ROSAT PSPC count rate maps of A3667. The
simulated PSPC maps were generated in the same energy band (0.4-2.0
keV), with the same pixel size ($15\arcsec\times15\arcsec$) and with
the same aim point ($20^h12^m31.20^s$, $-56\arcdeg 49\arcmin
12.0\arcsec$) as the real observation
(Section~\ref{sec:viper-data}). However, rather than covering the full
$512\times512$ pixel area of the ESAS map, the simulated map covered
only $256\times256$ pixels (roughly $1^{\circ}\times1^{\circ}$).  The
eight inputs to the ROSAT PSPC simulation were as follows;

\begin{itemize}

\item{$\beta$, $\theta_c$ and $\varepsilon$ to define the shape of
ellipsoidal $\beta$-model;}

\item{$\phi$ to define the projected inclination angle of
cluster. This was defined so that $\phi=0^{\circ}$ when the major axis
points toward the West and $\phi=90^{\circ}$ when the major axis
points toward the North;}

\item{$\alpha_0$ and $\delta_0$ to define the sky coordinate of the
center of the simulated cluster. These parameters were free to vary
and so the center of the simulated map did not necessary coincide with
the geometric center of the map;}

\item{$C_{\rm X 0}$ to define the 0.4-2.0 keV ROSAT PSPC count rate in
a $15\arcsec\times15\arcsec$ pixel located at
($\alpha_0$,$\delta_0$). We note that this is the central count rate
{\it before} the correction for the instrument point spread function
(PSF) had been applied (see below);}

\item {$C_{\rm b}$ to define a constant X-ray background
in the same units as $C_{\rm X 0}$.}

\end{itemize}

Before embarking on a maximum likelihood analysis, the simulated
images had to be corrected for the effects of the PSPC PSF.  This
correction was complicated by the fact that the PSF is a function of
the radial distance from the center of the field of view. To account
for the changing PSF, we convolved each simulated image with 10
different smoothing kernels. Each kernel corresponded to a model PSF
appropriate to a different, evenly spaced, radius from the aim
point. The ten resulting convolved images were then co-added using the
weighting scheme outlined in Equation~\ref{eqn:weights}. This method
resulted in a dramatic computational speed-up, compared to a direct
calculation of a convolution with a smoothly varying PSF model,
because it allowed us to use Fast Fourier Transforms (see
Appendix~\ref{app:point} for more information). Even so, it would have
been computationally intensive to convolve the entire $256\times256$
area of the simulated maps, so instead, we convolved only those pixels
with a simulated count rate of least 2\% of $C_{\rm X 0}$. The number
of pixels meeting that criterion varied depending on the values of the
input parameters, but typically only comprised those within a $\simeq60$
pixel radius of the center.  This simplification was justified by the
fact that cluster profiles vary slowly at large radii, meaning the
shape of the PSF has little impact on the per pixel count rate outside
the cluster core. We numerically tested how much impact PSF
convolution has on the outer regions of the cluster model and found it
to be insignificant. To create a final simulated image, the center of
the original $256\times256$ pixel simulation was replaced by the
circular convolved region.  We note that the edges of the convolved
region were cropped before insertion into the larger map in order to
correct for the effects of the artificial periodic boundary condition
imposed by the FFT.

\subsection{Simulated Corona Images}
\label{sec:sim-viper}

For every simulated ROSAT PSPC X-ray count rate image, we generate a
complementary simulated Corona brightness temperature image.  In the
non-relativistic and low frequency limit, the thermal SZE brightness
temperature can be calculated for a given line of sight using the
following equation;

\begin{equation} \frac{\Delta
T}{{T_{{\rm CMB}}}} = {Y}(x\,coth(x/2) - 4), \label{eqn:deltaT}
\end{equation} 

where $T_{{\rm CMB}}$ is the average temperature of the CMB (2.73K,
Fixen {\it et al.} 1996); $x$ is a dimensionless observing frequency,
$x=\frac{h\nu}{k {T_{{\rm CMB}} }}$; and the Compton-$Y$ parameter, is
given by the line of sight integral \begin{equation} {Y} = \sigma_T
\int_{-\infty}^\infty \frac{k T_e}{m_e c^2} n_e dl,
\label{eqn:compton-y} \end{equation} where $\sigma_T$ is the Thompson
cross section and $n_e$, $T_e$ and $m_e$ are the electron number
density, temperature and mass respectively.

Each simulated Corona image was generated with the same pixel size
($113\arcsec\times113\arcsec$) and assuming the same observing
frequency ($\nu=40 GHz$) as the raster scan image described above
(Section~\ref{sec:viper-data}). The same input values for $\beta$,
$\theta_c$, $\varepsilon$, $\alpha_0, \delta_0$ and $\phi$ were used
to generate both the Corona simulation and the companion ROSAT
simulation. In addition, input values for the electron temperature,
$T_e$, and the central electron number density, $n_{e0}$. were
required. The electron temperature enters the brightness temperature
calculations via the the Compton-$Y$ parameter
(Equation~\ref{eqn:compton-y}) and was assumed to be constant
throughout the cluster.  The central electron number density enters
via the normalization of the $\beta$-model
(Equation~\ref{eqn:beta-ne}) and was calculated using the expression
given in Equation~\ref{eqn:Ktone}, see section~\ref{sec:Ktone}.

The brightness temperature, $\Delta T$, is proportional to the line of
sight integral of the electron number density
(Equations~\ref{eqn:deltaT} \& \ref{eqn:compton-y}). If $n_e$ follows
a $\beta$-model (Equation~\ref{eqn:beta-ne}), one can obtain a simple
expression for $\Delta T$ using infinite integration limits that is
similar to the one (Equation~\ref{eqn:xray-sb}) we used to produce the
simulated ROSAT maps, {\it i.e.};

\begin{equation} \Delta T = \Delta T_0\left(1
+\frac{\theta_r^2}{\theta_c^2}\right)^{-3\beta+1/2}, \label{eqn:sze-sb}
\end{equation} 

where $\Delta T_0$ is the central brightness temperature and
$\theta_r$, $\theta_c$ and $D_A$ are as defined for
Equation~\ref{eqn:xray-sb}.  Unfortunately, as shown by Puy et
al. (2000), using infinite integration limits leads to a significant
error in the derived value of $H_0$. Therefore, rather than use
Equation~\ref{eqn:sze-sb}, we integrated Equation~\ref{eqn:beta-ne}
numerically over the range $-14 r_c< r<+14 r_c$. By using these limits
we avoided included spurious SZE signal at large radii. We note that
as a result of our definition of $r$ (Equation~\ref{eqn:ellipse}),
these integration limits defined an iso-density ellipsoid around the
cluster.

After the completion of the numerical integrations, the final steps
needed to produce a simulated Corona were to convolve the simulated
map with the Viper beam and then apply the same offset subtraction
scheme used on the real data (Section~\ref{sec:viper-data}).  We
performed the latter step because offset subtraction removes certain
linear modes from the measured SZE signal and these modes must also be
removed from the simulated images before the maximum likelihood
analysis takes place.

\subsubsection{Calculating the Central Electron Number Density}
\label{sec:Ktone}

In Appendix~\ref{app:ne0}, we outline the derivation of the following
expression for the central electron number density; \begin{equation}
n_{e0} = \sqrt{C_{\rm X 0} F} \sqrt{\frac{ 4\pi(1+z)^4
\varepsilon^{\zeta} \kappa 10^{14}}{D_A \theta_c \Omega }\frac{
\Gamma(3 \beta) }{\sqrt{\pi}\Gamma(3 \beta - 1/2)}}.
\label{eqn:Ktone} \end{equation}

This expression had to be evaluated before each simulated Corona map
was created. The majority of the parameters in the expression have
been introduced earlier, but others; $\kappa$, $\Omega$, $F$,
$\zeta$ and $D_A$, require additional explanation:  \begin{itemize}

\item $\kappa$ is the ratio of electrons to protons in the
intracluster medium. For our analysis, we used $\kappa=1.16$, which is
appropriate for a fully ionized plasma with 30\% solar abundances
(Feldman 1992).

\item $\Omega$ is the area over which the central X-ray count rate,
$C_{\rm X 0}$, is measured (or simulated). For our analysis we used a
square ROSAT PSPC pixel ($15\arcsec\times15\arcsec$);

\item $F$ is a scaling factor that converts an X-ray count rate into
an energy flux. The former can be measured with the ROSAT PSPC and the
latter can be calculated using theoretical emission models.  The value
of $F$ varies with electron temperature ($T_e$), redshift ($z$), metal
abundance ($Z$) and hydrogen column density ($n_H$). Assuming
isothermality, each cluster has a characteristic $F$ value, which can
be calculated using an X-ray spectral package such as XSPEC (Arnaud
1996). For our analysis of A3667 (Section~\ref{subsec:max-likefit}),
we allowed $T_e$ and $n_H$ to vary, but kept the redshift and metal
abundance fixed at $z=0.055$ and $Z=0.3Z_{\odot}$ respectively: the
redshift of the A3667 is well known (more than 100 constituent galaxy
redshifts have been measured, Katgert et al. 1998) and the ROSAT count
rate varies little with metal abundance (see Hughes \& Birkinshaw 1998
for a discussion of how little effect $Z$ has on measured values of
$H_0$). We created a look up table of $F$ factors covering a range of
$T_e$ and $n_H$ values using the {\tt raymond} (Raymond \& Smith
1976), {\tt wabs} (Morrison \& McCammon 1983) and {\tt fakeit}
commands in XSPEC, together with the appropriate ROSAT PSPC response
function.

\item $\zeta$ is the power to which we raise $\varepsilon$ to obtain
the stretch of the beta model in the direction of the line of sight
axis (Equation~\ref{eqn:ellipse}).  It is impossible to constrain both
$\zeta$ and $H_0$ from the projected X-ray and SZE data
alone. Following, Grainger (2001) we set $\zeta$ to one half, which
implies that the stretch along the line of sight is the geometric mean
of the stretch along the other two axes.  

\item $D_A$ is inversely proportion to the Hubble constant and is also
a function of acceleration parameter, $q_0$. During our analysis,
$H_0$ was free to vary, but $q_0$ was fixed at $q_0=0.5$.  We note
that at the redshift of A3667, the choice of $q_0$ has negligible
impact on our $H_0$ determination. This is in contrast to higher
redshift SZE observations; Hughes \& Birkinshaw (1998) claim that by
$z=0.5$, current uncertainties in the underlying cosmological model
result in a 10-20\% systematic error in $H_0$.

\end{itemize}

In summary, under the assumption that the electron number density
follows a $\beta$ profile, the central density, $n_{e0}$, can be
expressed as a function of the following parameters; $C_{\rm X 0}$,
$\theta_c$, $\beta$, $\varepsilon$, $\zeta$, $\kappa$, $z$, $D_A$
(hence $H_0$ and $q_0$) and $F$ (hence $n_H$, $z$, $Z$ and
$T_e$). Of these, $z$, $Z$, $\kappa$, $q_0$ and $\zeta$ were fixed
during the model fitting.

\subsection{Calculating the Joint Likelihood of the Simulated Maps}

\label{subsec:max-likefit}

After converting the simulated and observed count rate maps into units
of counts per pixel using the ESAS exposure map
(Section~\ref{sec:viper-data}), we calculated the log likelihood of
the simulated ROSAT PSPC images using the following expression, under
the assumption of Poisson statistics

\begin{equation} \ln{\cal L}_{\rm{\sc PSPC}} = \sum_i
-\ln(N_{\rm obs}(i)!) + N_{{\rm obs}}(i)\,\ln(N_{{\rm sim}}(i)) - N_{{\rm sim}}(i),  \label{eqn:logL-Xray} \end{equation}

where $N_{{\rm sim}}(i)$ is number of counts in the $i^{th}$ pixel of
the simulated map and $N_{{\rm obs}}(i)$ is the corresponding number
of observed counts. We carried out the summation over almost all
pixels within a $1\times1\arcdeg$ square centered on the PSPC aim
point.  The pixels we excluded from the sum had been flagged as
belonging to foreground or background X-ray point sources (see
Appendix~\ref{app:point}).

We also calculated the log likelihood function for the simulated
Corona images assuming Gaussian noise using the following expression

\begin{equation} \ln{\cal L}_{\rm{Corona}} = 
-\frac{1}{2}(\ln( 2\pi |\bar{\Sigma}|)
+ \Delta'\bar{\Sigma}^{-1}\Delta) \label{eqn:logL-SZ} \end{equation}


where the vector, $\Delta$, is defined as $\Delta= d_{\rm obs} -
d_{\rm sim}$; $d_{\rm obs}$ is a vector containing the measured
brightness temperature values ($\Delta T_{\rm obs}$) in each of the
pixels under test and $d_{\rm sim}$ is the corresponding vector of
$\Delta T_{\rm sim}$. In Equation~\ref{eqn:logL-SZ}, $\bar\Sigma$ is
the covariance matrix that describes the correlation between the
pixels under test.  If the pixels were uncorrelated, then the
covariance matrix would be populated by zeros except along the
diagonal, where it would be populated by the variance in the
corresponding pixel. In the case of the Corona map, the covariance
matrix is more complex because the pixels are correlated via
atmospheric noise, the finite response time of the detector and the
linear interpolation used during the map making process.  To evaluate
$\bar\Sigma$ we chose 28 pixels lying within one beam ($15.7\arcmin$
FWHM) of the center of the co-added, but unsmoothed, Corona map of
A3667.  We note that we decided to analyze data from this region,
rather than the from the whole Corona map, because only near the
center of the cluster did the SZE signal exceed the background
(statistical plus CMB) noise. Also we wished to avoid regions
overlapping with known radio sources. From the 1,518 individual raster
images described above (Section~\ref{sec:viper-data}), we compiled a
28 by 1,518 matrix ($D$) with a column for each of the chosen pixels
and a row containing the corresponding brightness temperature
measurement. Assuming that for any given pixel, the 1,518 measurements
were normally distributed, the covariance matrix for each row of $D$
was estimated as follows; $\Sigma = D^T C D$, where $D^T$ is the
transpose of $D$ and $C$ is the centering matrix. The latter is given
by $C = {\bf I} - \frac{1}{n} ({\bf 1}_n {\bf 1}^T_n)$. Here, ${\bf
I}$ is the identity matrix; ${\bf 1}_n$ is a column vector of ones of
length $n$ (in this case $n=1518$); and ${\bf 1}^T_n$ is the transpose
of ${\bf 1}_n$. Since the 1,518 scans are independent and the errors
are normally distributed, we can determine the covariance matrix for
vector $d_{\rm obs}$, $\bar\Sigma$, by dividing the covariance matrix
for each row of D by the number of scans; {\it i.e} $\bar \Sigma =
\Sigma/1518$.  We note that the 28 pixels chosen for the analysis
represented only a subset of the total number of pixels within the
central beam. The pixels were selected in a checker board fashion, to
avoid including any highly correlated nearest neighbor pairs in the
analysis. The checkerboard selection method ensured that the
covariance matrix was non-singular ({\it i.e.}  invertible) without
throwing away too much data.

The joint log likelihood of the two simulated maps, $ln {\cal L}_{\rm
joint}$, is given by the sum of the individual log likelihoods, {\it i.e.}

\begin{equation}
ln {\cal L}_{\rm joint}= ln {\cal L}_{\rm{PSPC}} + ln {\cal L}_{{\rm Corona}} + ln
{\cal L}_{T_e} + ln {\cal L}_{n_H},
\label{eqn:jointL}
\end{equation}

where $ln {\cal L}_{\rm{PSPC}}$ and $ln {\cal L}_{{\rm Corona}}$ were
defined above and $ln {\cal L}_{T_e}$ and $ln {\cal L}_{n_H}$ are the
log likelihoods of the electron temperature, $T_e$, and hydrogen
column density, $n_H$, values used to generate the simulated Corona
map.  The latter parameters were treated differently to the other free
parameters in our model because they have been measured by independent
experiments. Under the assumption of Gaussian statistics, their log
likelihood functions are given by

\begin{equation} ln {\cal L}_{T_e} = -ln(\sigma_{T_e}\sqrt{2\pi})- (T_{e}^{\rm in} - T_e^{\rm obs})^2/(2\sigma_{T_e}^2),
\label{eqn:logL-Te} \end{equation}
\begin{equation} ln {\cal L}_{n_H} = -ln(\sigma_{n_H}\sqrt{2\pi})- (n_H^{\rm in} - n_H^{\rm obs})^2/(2\sigma_{n_H}^2),
\label{eqn:logL-nH} \end{equation}

where $T_e^{\rm in}$ and $n_H^{\rm in}$ are the simulation input
values and $T_e^{\rm obs}$, $n_H^{\rm obs}$, $\sigma_{T_e}$ and
$\sigma_{n_H}$ are the measured values and their corresponding 1 sigma
errors. In our analysis we used $T_e^{\rm obs}=7.31$ keV;
$\sigma_{T_e}=0.17$ keV (Markevitch, Vikhlinin \& Murray 2001) and
$n_H^{\rm obs}= 4.71\times 10^{20}$ cm$^{-2}$; $\sigma_{n_H}=
1\times 10^{20}$ cm$^{-2}$ (Dickey \& Lockman 1990).

\subsubsection{Best fit model and errors on best fit parameters}

As described above, we were able to calculate a joint log likelihood
for each pair of simulated A3667 maps using
Equation~\ref{eqn:jointL}. The most likely ellipsoidal isothermal
$\beta$-model for A3667 was that which maximized the joint
likelihood. In order to find the most likely model we created several
thousand pairs of simulated images, adjusting the 11 free input
parameters ($C_{\rm X 0}$, $C_{\rm b}$, $\theta_c$, $\beta$,
$\varepsilon$, $\phi$, $\alpha_0$, $\delta_0$, $H_0$, $T_e$ and $n_H$)
slightly each time.  The parameter adjustments were governed by the
downhill simplex (Nelder and Mead 1965) routine.  We note that the
first term ($-\ln(N_{obs}(i)!)$) in Equation~\ref{eqn:logL-Xray} is
independent of the model being tested, and therefore to speed up the
calculations, was not evaluated during the $ln \cal L_{\rm joint}$
maximization.  The best fit values for nine of the eleven free
parameters in the fit are given in Table~\ref{tab:errors}. The best
fit values for the other two parameters, $n_H$ and $T_e$, were
identical (to 3 significant figures) to the values determined by
independent experiments (see above). The central electron density
corresponding to our best fit model (see Equation~\ref{eqn:Ktone}) is
$n_{e0}=4.05\times10^{-3}$ cm$^{-3}$.  

After the most likely $\beta$-model had been found, we used the
likelihood ratio test (LRT) to define errors on the various input
parameters in the fit. The application of the LRT to SZE analysis has
been described elsewhere ({\it e.g.} Hughes \& Birkinshaw 1998; Reese
et al. 2000), but we review the salient points here. The variable
${\cal S} \equiv - 2 {\rm ln}({\cal L})$ has a minimum ${\cal
S}_{min}$ for the best fit model ({\it i.e.} when ${\cal L}$ is at a
maximum). The Cash statistic (Cash 1979) is defined as ${\cal C}={\cal
S}-{\cal S}_{min}$ and is, therefore, equal to zero for the parameter
set that defines the best fit model and greater than zero for any
other set.  The 68\% confidence region for a particular input
parameter can be estimated by finding the range of values that it can
hold for which the Cash statistic that is less than unity. When making
the Cash statistic calculations, we hold the parameter at a certain
value and then carry out a likelihood maximization by varying the
other parameters in the model.  Several values of the parameter need
to be tested before the ${\cal C}<1$ region can be determined. The
traditional approach to the LRT involves measuring ${\cal L}$, and
hence ${\cal C}$, at evenly spaced grid points in the n-dimensional
parameter space. However, this approach is computationally prohibitive
for a model with 11 parameters.  We, therefore, devised a new LRT
method which was both more efficient and more objective then the
traditional method (it is more objective because it requires no a
priori knowledge of the parameter error distributions).

\begin{figure*} {\centering\leavevmode\epsfysize=6.in
\epsfbox{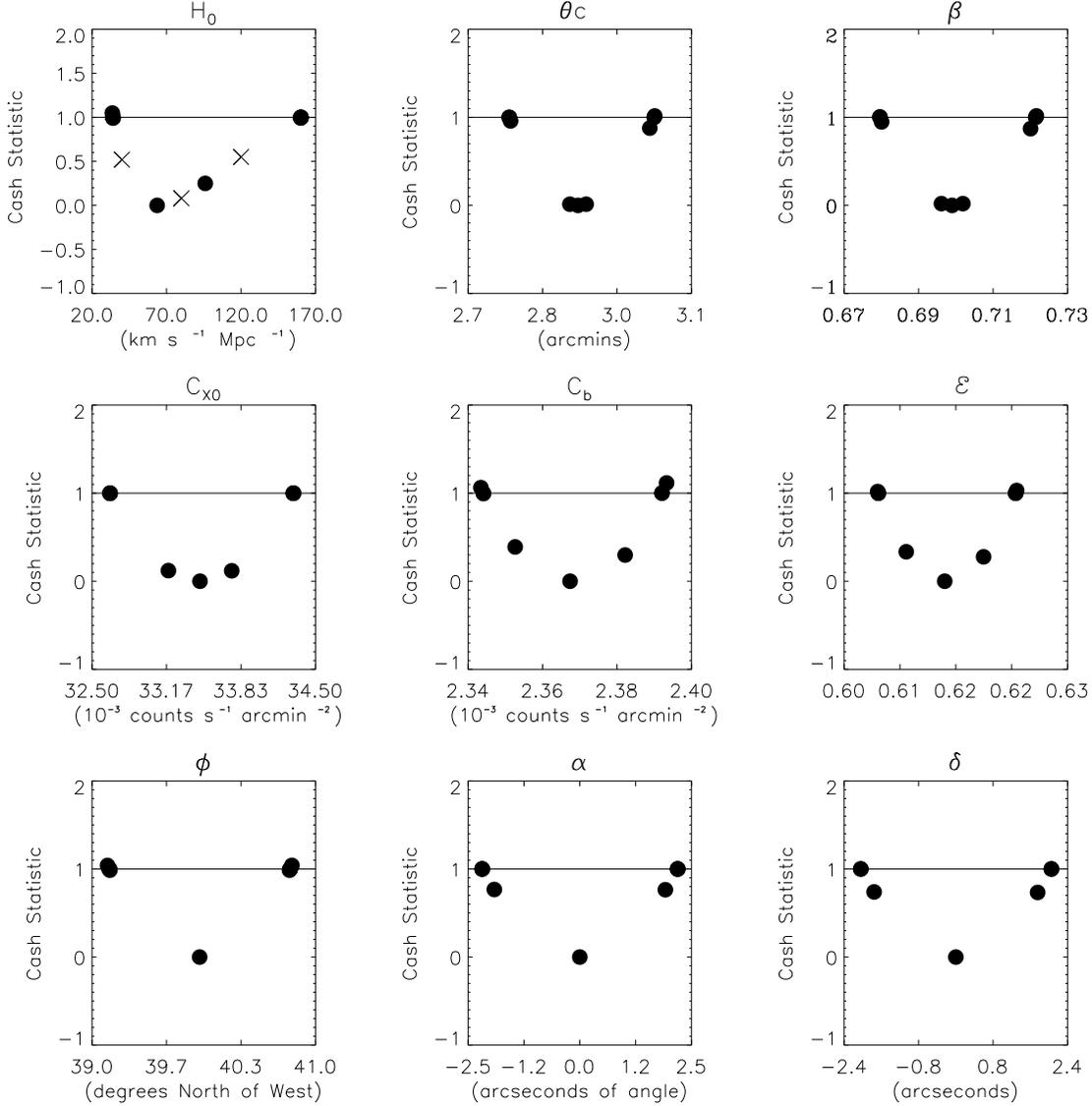}} \figcaption{\label{Fig:cashstats} The Cash
statistic as a function of: Hubble Constant (top left); core radius
(top center); $\beta$ (top left); central (0.5-2.0 keV) count rate
(center left); background (0.5-2.0 keV) count rate (center center);
ellipticity (center right); orientation angle (bottom left); central
right ascension coordinate (bottom center); central declination
coordinate (bottom right). The ${\cal C}=1$ crossing points define the
edges of the 68\% confidence region for each parameter. The black dots
indicate the parameter values chosen by our automated error analysis
software; notice how few values need to be tested before the ${\cal
C}=1$ crossing points were found. For $H_0$ we added three extra test
values by hand to demonstrate that there were not additional minima in
the ${\cal C}(H_0)$ function.}  \end{figure*}

Our method proceeded as follows. First we did a rapid search to find
values of the parameter under test that yielded Cash statistic values
greater than unity after marginalizing over the other 10 free
parameters.  We initiated this search by fixing the parameter at a
point 50\% above its best fit value. We then performed a likelihood
maximization over the remaining 10 free parameters. We repeated this
for a fixed point 50\% below the best fit value. If these $\pm 50\%$
points did not yield ${\cal C}>1$, then we repeatedly doubled the
distance between the tested values until they did. In practice, we
only had to use this doubling technique for the Hubble Constant ({\it
i.e.} all the other parameters had $1 \sigma$ errors that were less
than 50\%). After finding two locations where ${\cal C}>1$, and
knowing the location of the ${\cal C}=0$ point, we estimated the
locations of the ${\cal C}=1$ points using a third order beta spline
fit. The Cash statistic was re-evaluated at the estimated ${\cal C}=1$
crossing points and the spline fit performed again. The process was
repeated two more times. In this way we built up at least nine $\cal
C$ measurements; {\it i.e.} one at, four above and four below the best
fit value. With at least nine inputs to the spline fit, it was
possible to accurately determine of the ${\cal C}=1$ crossing
points. We performed the LRT error analysis for nine of the
eleven\footnote{For the other two, $n_H$ and $T_e$, error estimates
were available from independent experiments, see
section~\ref{subsec:max-likefit}.} free parameters in our model fit,
see Figure~\ref{Fig:cashstats} and Table~\ref{tab:errors}.  In general
the the errors were roughly quadratic, as expected for Gaussian
errors; however, in the case of $H_0$, the error distribution had a
broad tail on the positive side. To demonstrate that the Cash
Statistic was a smooth function of $H_0$, we tested some additional
$H_0$ values there were not selected by the root finding algorithm
(represented by the crosses on Figure~\ref{Fig:cashstats}).  Our best
fit value for the Hubble Constant is $64^{+96}_{-30}$ km s$^{-1}$
Mpc$^{-1}$ (68\% confidence limits), see below for a discussion.  Best
fit values for the other parameters can be found in
Table~\ref{tab:errors}.  For completeness, we list here published core
radius and $\beta$ values derived from previous fits to the ROSAT PSPC
observation of A3667: $\theta=2.86\arcmin$, $\beta=0.54$ (Mohr,
Mathiesen \& Evrard 1999); $\theta_c=3.54^{+1.4}_{-1.1}\arcmin$,
$\beta=0.67^{+0.19}_{-0.12}$ (Neumann \& Arnuad (1999; model B);
$\theta=4.29\pm0.96\arcmin$, $\beta=0.589\pm0.051$ (Mason \& Myers
2000).  Despite the fact that the previous fits were made to average
surface brightness profiles, rather than to the image directly, the
published values are generally consistent with those in
Table~\ref{tab:errors}.

The error on our $H_0$ measurement is primarily due to the statistical
instrument noise in the Corona map; we have $\simeq40\%$ error on the
measurement of the central temperature decrement and  $H_0\propto\Delta
T^2$. Other sources of random noise are less important because we have
access to high quality X-ray data from ROSAT and Chandra. Our best fit
$H_0$ value compares favorably with other recent SZE measurements;
$H_0=60^{+4}_{-4}(^{+13}_{-18})$ km s$^{-1}$ Mpc$^{-1}$ (Reese et
al. 2002; systematic errors appear in parentheses),
$H_0=66^{+14}_{-11}(^{+5}_{-5})$ km s$^{-1}$ Mpc$^{-1}$ (Mason, Myers
\& Readhead 2001), $H_0=H_0=65^{+8}_{-7}$ km s$^{-1}$ Mpc$^{-1}$
(Jones et al. 2001).  Unlike our analysis, these $H_0$ measurements
were based on observations of multiple clusters (18, 7 and 5 clusters
respectively). We note that errors of $\simeq\pm30$\% and exceeding
$\pm50$\% are quoted in Jones et al. (2001) sample and Mason, Myers \&
Readhead (2001) respectively on $H_0$ measurements derived from single
clusters, {\it e.g.}  $H_0=67^{+62}_{-28}$ km s$^{-1}$ Mpc$^{-1}$ for
the A2256 $H_0$ measurement in Mason, Myers \& Readhead (2001).  These
values demonstrates that, even though the errors on our $H_0$
measurement are large, they are not atypical, especially when compared
to other single dish experiments.

\begin{table*}
\caption{\label{tab:errors}The results of the joint maximum likelihood fit with 68\% confidence limits}
\small{
\begin{tabular}
{cccccccc} 

$\theta_c$ 
&$\beta$
&$C_{\rm X0}^1$ 
&$C_{\rm b}^1$
&$\varepsilon$ 
&$\phi$
&\multicolumn{2}{c}{$\alpha_0$ (J2000) $\delta_0$}\\ \hline

$2.9\pm{+0.1}\arcmin$ 

&$0.7{\pm 0.02}$ 

& $33.4\pm0.8$ 

& $2.37{\pm 0.02}$ 
 
&$0.614\pm 0.009$ 

&$40.0{\pm 0.8}\arcdeg$	

&$20^{h}12^{m} 28.9 \pm 0.1^s$ 

&$-56\arcdeg 49\arcmin 51 \pm 2\arcsec$ 


\end{tabular} } Notes: 1: Count rates in the 0.4-2.0 keV band in
units of $\times 10^{-3}$ counts $s^{-1}$ arcmin$^{-2}$

\end{table*}

\section{Conclusions}

\label{sec:conclusions}

As the first stage of the Viper Sunyaev-Zel'dovich Survey (VSZS), we
have presented a 40 GHz raster scan map of the region surrounding the
$z=0.055$ cluster A3667. The cluster was observed during the Antarctic
winter of 1999 using the Corona instrument ($15.7\arcmin$ FWHM beam)
on the Viper Telescope at the South Pole.  The Corona image of A3667
is one of the first Sunyaev-Zel'dovich effect (SZE) images of a low
redshift cluster.  Currently, only the Viper telescope and the Cosmic
Background Imager interferometer (Udomprasert, Mason \& Readhead 2001)
are able to make SZE images of low redshift clusters. One of the
advantages of low redshift observations of the SZE is that
complimentary data at other wavelengths, for example those necessary
to make Hubble Constant estimates, are comparatively easy to
obtain. The Corona image of A3367 is also one of the first direct
({\it i.e.}  rather than interferometer) SZE images; only two other
clusters have published direct SZE images.

We have used the 40GHz map of A3667 in conjunction with a deep ROSAT
PSPC (X-ray) image to make a maximum likelihood fit to an isothermal
tri-axial ellipsoidal $\beta$ model. We have determined 68\%
confidence regions for nine of the eleven free parameters in the model
using a modified Likelihood Ratio Test. Our analysis method includes
new approaches to point source detection in X-ray images with varying
point spread functions and also to error estimation in
multi-dimensional parameter space. These innovations will be applied
during the analysis of other VSZS clusters, but are also relevant to
other SZE experiments. Our measurement of the Hubble Constant, $H_0=
64^{+96}_{-30}$ km s$^{-1}$ Mpc$^{-1}$, is in good agreement with
previous, SZE determined, measurements. The error on this measurement
is large, but not atypical for single dish experiments, and is being
driven by instrument noise in the Corona map. Other parameters in the
fit have been determined to much higher precision (1-3\%).

The Corona instrument was retired at the end of 2000 and replaced with
the ACBAR instrument (Runyan et al. 2002). ACBAR offers both improved
spatial resolution ($5\arcmin$ FWHM beam) compared to Corona and
simultaneous multi-wavelength (150, 220, 280 GHz) capabilities.  Using
ACBAR, in combination with data from CTIO, XMM-Newton and Chandra, we
have continued the VSZS and are working toward completion of the SZE,
X-ray and weak lensing observations of a complete, luminosity limited,
sample of $\simeq10$ clusters.  These observations will improve our
$H_0$ measurement in several ways. First, long duration ACBAR
observations produce more sensitive maps than was possible with Corona
(see Kuo et al. 2002). Second, we can minimize the contribution of
primary CMB anisotropies in our SZE images by combining data at the
three ACBAR observing frequencies (Gomez et al. 2002). This feature
becomes important when the instrument noise gets down to the level of
the CMB noise and it is noteworthy that most other SZE experiments
only operate at a single frequency; Diabolo (Pointecouteau 1999, 2001
\& 2002), BOLOCAM (Mauskopf et al. 2000b) and SuZIE (Holzapfel 1997
a\&b) are exceptions, but these are not optimized to image low
redshift clusters. Third, combining observations of several clusters
drawn from a statistically complete sample reduces both random errors
and systematic errors associated with orientation bias ({\it e.g.}
Grainger 2001). Finally, the improved spatial resolution of ACBAR over
Corona, means that we will be able to use independent beams to probe
different lines of the sight through each cluster. We only used data
from the central beam of the A3667 Corona image for the analysis
above, but by using multiple lines of sight we can reduce the error on
the overall measurement. Together, these improvements will allow us to
measure the Hubble constant with significantly higher precision than
was possible with the Corona observation of A3667 presented herein.

\paragraph{Acknowledgments}

We thank Eric Reese, Jack Hughes and Brian Mason for useful advice
concerning the X-ray analysis and Jullian Borril for assistance with
the calculation of the CMB noise in the Corona map. We thank the SHARC
collaboration for allowing us to use their reduced PSPC map of A3667
and Dick Hunstead and collaborators for supplying the MOST map of
A3367.  AKR, CMC, JBP and PG acknowledge financial support from the
NASA LTSA program for grant NAG5-7926, ``The Viper Sunyaev-Zel'dovich
Survey''. CMC was the recipient of CMU summer research grants for
undergraduates during 1998 and 1999. Computer equipment for this
project was purchased using a AAS small research grant.  Additional
computing resources were kindly made available to us by the PICA group
(picagroup.org). Finally we thank the ACBAR collaboration of their
continued support of low redshift cluster observations with Viper.


\appendix

\section{Derivation of the Equation for the Central Electron Number Density}

\label{app:ne0} We detail below how we arrived at
Equation~\ref{eqn:Ktone}. We begin with a definition for the the
normalization, $K$, of the XSPEC Raymond-Smith model (Arnaud et al. 1996);

\begin{equation}
{K} = \frac{10^{-14}}{4 \pi D_L^2} \int n_e n_p dV.
\label{eqn:K}
\end{equation}

Here $n_e$ and $n_p$ are the electron and proton number densities
respectively and $D_L$ is the luminosity distance to the cluster. If
we define $\kappa$ to be the ratio of electrons to protons
substituting $D_L= (1+z)^2 D_A$, we can rewrite Equation~\ref{eqn:K}
as follows \begin{equation} {K} = \frac{10^{-14}}{4 \pi D_A^2
(1+z)^4\kappa} \int n_e^2 dV.  \label{eqn:K2} \end{equation}
Expressing the volume element as $dV = D_A^3 \Omega d\theta_l$, we can
rewrite Equation~\ref{eqn:K2} in terms of an integral along the line
of sight, \begin{equation} {K} = \frac{10^{-14} D_A \Omega}{4\pi
(1+z)^4\kappa} \int n_e^2 d\theta_l, \label{eqn:K3} \end{equation}

where $\Omega$ is a solid angle and where the distance along the line
of sight is expressed as an angle ($\theta_l=l/D_A$).

If we assume that the electron number density follows an ellipsoidal
$\beta$-model, we can write down an
expression for ${K}_0$, the Raymond-Smith model normalization at
the center of the cluster as follows: at the projected cluster
centroid the distance simplifies to $r=\varepsilon^\zeta l$
(since $x=y=0$, see Equation~\ref{eqn:ellipse}). Thus, along the line
of sight intersecting the cluster center, the electron number density
is given by

\begin{equation} 
n_e = n_{e0}\left(1+ \left(\frac{\varepsilon^\zeta l}{r_c}\right)^2\right)^{-3/2\beta},
\nonumber
\end{equation}

which is equivalent to the following if we express $l$ and $r_c$ as angles;
\begin{equation} 
n_e = n_{e0}\left(1+ \left(\frac{\varepsilon^\zeta \theta_l}{\theta_c}\right)^2\right)^{-3/2\beta}.
\label{eqn:ne02} 
\end{equation}

Combining Equations~\ref{eqn:K3} and \ref{eqn:ne02}, yields

\begin{equation} { K}_0 = \frac{10^{-14} D_A \Omega}{4\pi
(1+z)^4\kappa} \int \left(n_{e0}\left(1+\left(\frac{\varepsilon^\zeta
\theta_l}{\theta_c}\right)^2\right)^{-3/2\beta}\right)^2 d\theta_l,
\label{eqn:K4} \end{equation} 

which can be integrated using the following  identity\footnote{Taken from Mason 2000, Equation~C.3}  and substitution;\
\begin{eqnarray} 
\int_{-\infty}^{\infty}(1+A^2+B^2)^{-C} dA &=& \sqrt{\pi}\frac{\Gamma
(C-1/2)}{\Gamma (C)}(1+B^2)^{-C + 1/2} 
\label{eqn:identity} \\
u &=& \frac{\varepsilon^\zeta\theta_l}{\theta_c} \Rightarrow d\theta_l =
\frac{\theta_c}{\varepsilon^\zeta} du\,\,,
\end{eqnarray}

as follows;

\begin{eqnarray} 
 {K}_0
&=&\frac{10^{-14} D_A \Omega n_{e0}^2\theta_c}{4\pi
(1+z)^4\kappa\varepsilon^\zeta} \int (1+u^2)^{-3\beta} du\\
{K}_0 &=& \frac{10^{-14} D_A \Omega n_{e0}^2\theta_c}{4\sqrt{\pi}
(1+z)^4\kappa\varepsilon^\zeta} \frac{\Gamma(3\beta -
1/2)}{\Gamma(3\beta)}.\label{eqn:K5}
\end{eqnarray}

We can determine the scaling between $K$, the normalization of the
Raymond-Smith model, and a ROSAT PSPC count rate for a partcular set
of input parameters ($T_e$, $Z$, $z$ and $n_H$) using XSPEC, see
Section~\ref{sec:Ktone}.  Defining, $F$, to be the scaling factor
between $K$ and an observed (or simulated) X-ray count rate ($C_{\rm
X}$) in a pixel of size $\Omega$ (see section~\ref{sec:Ktone}), {\it
i.e.}

\begin{equation} F=\frac{{K}}{C_{\rm X}}, 
\label{eqn:Fo}
\end{equation}

we finally arrive at Equation~\ref{eqn:Ktone};

\begin{equation} 
n_{e0} = \sqrt{C_{\rm X 0} F} \sqrt{\frac{ 4\pi(1+z)^4 \varepsilon^{\zeta}
\kappa 10^{14}}{D_A \theta_c \Omega }\frac{ \Gamma(3 \beta)}{\sqrt{\pi}\Gamma(3 \beta - 1/2)}}.
\end{equation} 

\section{Removing Point Sources from ROSAT PSPC Images}
\label{app:point}

We have devised a method based on Mexican hat wavelet convolutions to
identify point sources in X-ray images with variable point spread
functions (PSFs). This method is able to exploit the speed of the fast
Fourier transform (FFT) by approximating a smoothly varying PSF with
$n$ discrete smoothing kernels. This requires $O(n p \ln(p))$
operations, where $p$ is the number of pixels in the image, compared
to the $O(p^2)$ operations required for the exact convolution.  As $n$
approaches $\sqrt{p}$ the method becomes exact, but, in practice, $n$
can be much less than $\sqrt{p}$. We have used this approximation both
to find point sources in the ROSAT PSPC image of A3667, see below, and
to create simulated PSPC images, see Section~\ref{sec:sim-pspc}.  Our
technique could equally well be applied to any PSPC image and, after
small modifications, it could also be used to analyze images from
other X-ray detectors.  The authors have applied it successfully to
ROSAT HRI and XMM images.

The method we use to identify the point sources draws on the
theoretical framework set up in Damiani et al. (1997, D97 hereafter)
and has two parts: the selection of the optimum wavelets and the
identification of sources. A brief summary is as follows. First the
optimum Mexican hat wavelets to extract point sources located at a set
of $n$ radial distances from the center of the PSPC FOV are selected.
The X-ray image under analysis is then convolved with each of these
wavelets. Next, a linear combination of the $n$ convolution products
is generated. The resulting combined map approximates a map that has
been generated by convolving the X-ray image with a smoothly varying
kernel (where that varying kernal describes the optimum wavelet at
every position in the image). We note that the optimum wavelets are
scaled so that when they are convolved with an X-ray image the result
is in units of point source photon counts (see D97). For the analysis
of the PSPC observation of A3667 described above
(Section~\ref{subsec:max-likefit}), we used $n = 10$ different
wavelets corresponding to 10 evenly spaced distances from the center
of the PSPC FOV.

The Mexican hat wavelet used as our smoothing kernal is defined by:

\begin{eqnarray}
z(x,y,a) &=& \frac{\sqrt{x^2 + y^2}}{a} \nonumber \\
g(x,y,a) \equiv g(z(x,y,a)) &=& (2 - z^2) e^{-z^2/2},
\label{Equation:mexhat}
\end{eqnarray}

where, $a$ is the so called scale parameter.  Using an optimization
procedure, we can find the kernal best suited to find point sources
at a particular radius, {\it i.e.} the value of $a$ that maximizes the
following expression:

 \begin{equation}
\label{eqn:costFunction}
c(a) = \sum_i \sum_j \frac{g(i-i_0, j-j_0, a) s(i,j)}{a},
\end{equation}
where $s(i,j)$ is the simulated point source image positioned at
the center, $(i_0,j_0)$, of a large array and
$s$ is normalized so that $\sum_i \sum_j s(i,j) = 1$.  We carry out
this optimization for a set of $n$ distances, $r_k$, from the center
of the PSPC FOV.  For each of the chosen distances we maximize $c(a)$
by varying $a$ using the Nelder-Mead downhill simplex algorithm and
solve for $a_k$, where $k \in \{0, 1, ..., n-1\}$.

To generate the simulated point source images, we used the empirically
determined PSPC PSF model of Hasinger et al. (1993). For any given
point source, this model is a function of both its photon energy
spectrum and its radial distance. For our analysis, we made the
simplifying assumption that all point sources can be described by a
generic absorbed AGN spectrum ($n_H=4\times10^{20}$ cm$^{-2}$,
$\alpha=-1$), since AGN's comprise the majority of the ROSAT point
source population (Mittaz et al. 1999).

After determining the optimal wavelet for a set of distances from the
center of the PSPC FOV, we scaled the wavelets so that the magnitude of
point source shaped objects was preserved in the convolution.  The
scaled Mexican hat wavelet that we use is for each of our $k$ chosen
distances is \begin{equation} g_k(i,j) = \frac{1}{a_k c(a_k)}
g(i,j,a_k). \label{eqn:scaling} \end{equation}
 
Next, using FFT's, we convolved the X-ray image under analysis 
with each of the $n$ optimum wavelets:
\begin{equation}
\rho_k(i,j) = \sum_l \sum_m g_k(l-i,m-j) v(l,m),
\label{eqn:waveconv}
\end{equation}
where $v(i,j)$ describes the X-ray photon count rate image.  
The convolved maps $\rho_k$ are then combined together:
\begin{equation} 
\hat{\rho}(i,j) = \sum_{k=0}^{n-1} h_k(i,j) \rho_k(i,j)
\label{eqn:weight-scheme}
\end{equation}
where the weights $h_k$ are defined as follows

\begin{eqnarray}
r &=& \sqrt{(i-i_0)^2 + (j-j_0)^2} \nonumber \\
h_k(i,j) \equiv h_k(r) &=&
\left\{\begin{array}{ccccc}
0 & {\rm if} & r < r_{k-1}   &             &\\
\frac{r-r_{k-1}}{r_k-r_{k-1}}   & {\rm if} & r_{k-1} \le r & \&  & r < r_k 
\\
\frac{r_{k+1} - r}{r_{k+1} - r_{k}} & {\rm if} & r_k \le r & \& & r < 
r_{k+1} \\
0 & {\rm if} & r > r_{k+1}   &             &
                \end{array}\right.
\label{eqn:weights}
\end{eqnarray}

and $(i_0,j_0)$ is the center of $v(i,j)$.  By virtue of the wavelet
scaling (equation~\ref{eqn:scaling}), the units of the combined map
are such that if a point source was centered on position $(i,j)$ then
$\hat{\rho}(i,j)$ would be an estimate of the photon count rate from
that point source. This count rate can then be converted into an
estimate of the number of photons collected from a point source, by
multiplication with an appropriate exposure map ($E(i,j)$).
Therefore, to excise point sources from an X-ray image, we can simply
apply a photon count threshold to $\hat{\rho}(i,j)\times E(i,j)$.

For the A3667 analysis described in section~\ref{subsec:max-likefit}
we excluded pixels above an 65 count threshold in
$\hat{\rho}(i,j)\times E(i,j)$ from the maximum likelihood analysis.
We also excluded pixels within a 4 pixels radius of
$\hat{\rho}(i,j)\times E(i,j)>65$ count regions. We note
that this 80 count threshold was chosen to meet the specific needs of
the analysis described in section~\ref{subsec:max-likefit}. A
different type of analysis, a different cluster observation, or a
different instrument would likely have led to a different choice.

\newpage

\newpage

\newpage
\begin{figure*}
{\centering\leavevmode\epsfysize=8.in \epsfbox{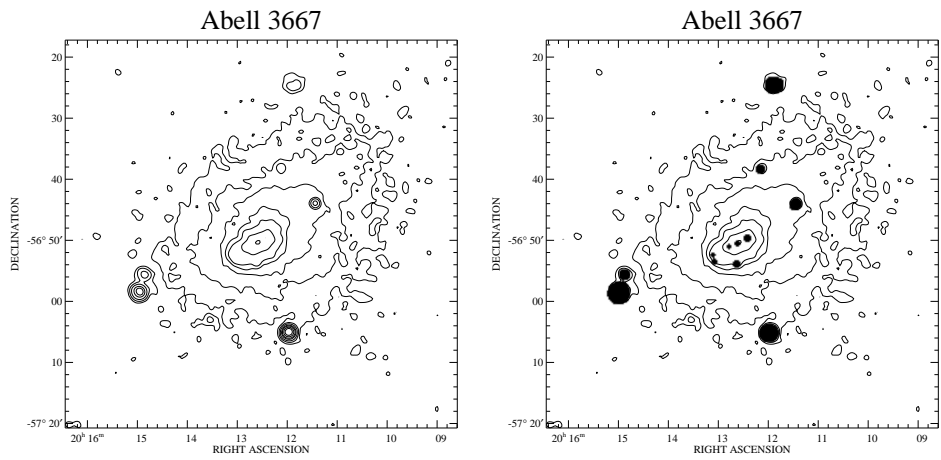}}
\vspace{-5in}
\figcaption{\label{Fig:PSPC}  
The 0.4-2.0 keV PSPC count rate image of A3667 (right) and the same
image after (right) point sources have been masked out. The masked
regions were not included in the maximum likelihood analysis.}
\end{figure*}

\newpage

\end{document}